# Amine-Linked Single Molecule Circuits: Systematic Trends Across Molecular Families

Mark S. Hybertsen[1,a], Latha Venkataraman[2,4,b], Jennifer E. Klare[3,4,c], Adam C. Whalley[3,4], Michael L. Steigerwald[3], and Colin Nuckolls[3,4]

[1]Center for Functional Nanomaterials, Brookhaven National Laboratory, Upton, New York

[2]Department of Applied Physics & Applied Mathematics, Columbia University, New York, New York

**[3]**Department of Chemistry, Columbia University, New York, New York

[4]Center for Electron Transport in Molecular Nanostructures, Columbia University, New York, New York

**Abstract.** A comprehensive review is presented of single molecule junction conductance measurements across families of molecules measured while breaking a gold point contact in a solution of molecules with amine end groups. A theoretical framework unifies the picture for the amine-gold link bonding and the tunnel coupling through the junction using Density Functional Theory based calculations. The reproducible electrical characteristics and utility for many molecules is shown to result from the selective binding between the gold electrodes and amine link groups through a donor-acceptor bond to undercoordinated gold atoms. While the bond energy is modest, the maximum force sustained by the junction is comparable to, but less than, that required to break gold point contacts. The calculated tunnel coupling provides conductance trends for all 41 molecule measurements presented here, as well as insight into the variability of conductance due to the conformational changes within molecules with torsional degrees of freedom. The calculated trends agree to within a factor of two of the measured values for conductance ranging from $10^{-7}\,G_0$ to $10^{-2}\,G_0$, where $G_0$ is the quantum of conductance ($2e^2/h$).

---

[a] mhyberts@bnl.gov
[b] lv2117@columbia.edu
[c] Present address: Department of Chemistry, Northwestern University, Evanston, Il.

# 1. Introduction

Understanding the transport characteristics of molecules bonded between metal electrodes is of fundamental interest with potential for impact in such areas as nanoscale electronic components [1]. It is generally expected that such transport characteristics are controlled by the intrinsic properties of the molecules, including their length, conformation, the gap between the highest occupied molecular orbital (HOMO) and the lowest unoccupied molecular orbital (LUMO) and the alignment of this gap to the metal Fermi level. Indeed, the concept of utilizing the flexibility of organic synthesis to design versatile molecular circuits depends on understanding the molecular structure, electronic transport relationships. However, experimental studies to systematically map such "design rules" have been difficult in practice. Broad variations in measured conductance through molecules linked to the electrodes via the widely used thiol chemistry have certainly hindered this process [2]. This may be attributable to intrinsic changes in electronic coupling through the thiol link for different local link configurations [3]. Experiments are performed using diverse platforms to probe either the conductance of one or a few molecules forming a circuit [4-14] or to probe a small patch of molecules forming an interface in an ensemble measurement [15-17]. While progress has been made in comparing different approaches [18], significant puzzles remain, for instance the substantial differences that have been observed between ensemble and single molecule measurements for the same species [19].

Some specific trends in molecular conductance have been measured using thiol linkages. The current-voltage characteristics of alkanes were measured with several approaches showing clear consistency with a non-resonant tunneling mechanism and the expected exponential dependence on molecule length [8, 20-22]. Measurements of the conductance through oligophenylenes and carotanoids of different length also showed exponential dependence on molecule length, but with a smaller decay constant as expected [23, 24]. Comparison of the conductance through alkanes to that through prototypical molecular wires with extended pi electron states (oligophenyleneethynylene, OPV and oligophenylenevinylene, OPV) showed substantially higher conductance through the conjugated molecules and a rational dependence on the HOMO-LUMO gap [23, 25-27]. In a series of asymmetric junctions formed with oligophenyl and acene monothiols, systematic changes in conductance with electrode Fermi level alignment were observed [28]. Comparison of benzenedimethanethiol to benzenedithiol shows lower conductance due to interpolating the non-conjugated methylene groups [29]. The histograms of alkanedithiol conductance measured by repeated reformation of the junction in solution have shown a few distinct families of peaks, suggestive of distinct and common local configurations of the junctions [22, 27, 30]. Probing the dependence on internal molecular conformation has been more difficult. Changes induced by optically induced switching have been observed [13, 31, ]. Finally, clear signatures for vibrational excitation of the molecule during electron transport have been observed [32, 33].

We have found that a new approach to single molecule circuit formation, the use of amine linkages with gold electrodes, has enabled us to perform reliable measurements of the conductance for several series of different molecules, both conjugated and saturated [34-37]. The experiments are performed by repeated formation and rupture of an Au-Au point contact in a solution of the target molecules [8]. Distinctive peaks emerge in the histograms of junction conductance collected from $O(10^4)$ cycles without further data selection. The data point to clear values for the most probable junction conductance for each diamino coupled molecule. The studies of different molecular families have probed length dependence, internal molecular conformation, the impact of substituents on the alignment of frontier orbital energies to the electrode Fermi energies and the HOMO-LUMO gap of aromatic molecules.

We have hypothesized that the filled lone pair on the amine group forms a donor-acceptor bond to a specific, coordinatively unsaturated Au atom in the junction [34]. Such atomic sites should be readily available in the as-formed junctions, following rupture of the Au-Au point contact. The selectivity and other characteristics of the donor-acceptor bonding have been supported by extensive Density Functional Theory (DFT) based calculations [34, 38]. In particular, this model for the link bonding gives an intuitively clear explanation for the reproducible electronic coupling through the link. The frontier orbitals on the link are dominated by the coupling of the lone pair on the N to the isotropic s-orbital on the Au atom to which the N bonds. This coupling follows the N-Au bond orientation with only small variations for the different local configurations of the junction probed over the course of a large number of measurements. A competing hypothesis would have been formation of an amide bond following loss of one of the hydrogens from the amine group. However, we have recently shown that methyl sulfide and dimethyl phosphine link groups also form reproducible single molecule circuits with alkanes [39]. This data was also clearly explained by the donor-acceptor bonding hypothesis and the removal of a methyl group during junction formation is highly unlikely.

Because the amine link scheme has proved successful for several distinct molecular families, we can start to make systematic comparisons of conductance between molecular families. In this paper, we provide a comparison of the experiments we have performed with amine linkages, extending the experiments for several families of molecules and providing new data for a set of short "molecular wires." The DFT based approach to model the donor-acceptor bonding and the junction properties are presented in a unified framework. This gives a picture of the mechanical properties of the junctions as well as a model with which we can compare the tunneling probability for the different amine linked junctions. In particular, the DFT based approach is shown to explain the main trends in the tunneling conductance across the entire data set of 41 distinct molecules. Since the calculated tunnel coupling is intimately connected with the DFT derived orbital energies, questions concerning intrinsic errors in comparison to spectroscopic energies (e.g. the ionization potential of the molecule) remain to be addressed [38].

The paper is organized with the experimental approach and data analysis methods described in Sect. 2. The theoretical methods are presented in Sect. 3, and our approach to model the bonding and conductance of amine linked junctions is developed in Sect. 4. In Sect. 5, we discuss three trends: length dependence comparing alkanes, oligophenyls and acenes; molecular wires and the role of conformation; comparison of tunnel coupling to measured conductance across five molecular families. Concluding remarks appear in Sect. 6.

## 2. Experimental Methods

We measure the conductance of metal-molecule-metal junctions by repeatedly forming and breaking Au point-contacts in a solution of the molecules using a home-built, simplified scanning tunneling microscope (STM) [8]. The Au point-contacts are formed as an Au wire tip is moved in and out of contact with an Au substrate. Conductance traces measured as a function of increasing tip-sample displacement reveal quantized conductance steps observed at multiples of $G_0$ ($2e^2/h$), the fundamental quantum of conductance. The steps indicate that the cross-section of the contact is reduced down to that of a few and eventually a single atomic chain [40]. When the single atom chain, with a conductance close to $G_0$ is broken in the absence of molecules, the conductance either decreases exponentially with the electrode displacement due to tunneling across the gap between the two Au electrodes or drops from $G_0$ to below our experimental detection limit, possibly due to the broken ends snapping back as the contacts relax. When the gold point contacts are broken in a solution of molecules, frequently, a molecule is trapped between the broken gold point-contacts, forming a metal-molecule-metal junction. The conductance of these junctions remains approximately constant as the junction is pulled apart. This results in additional steps seen in the conductance traces below $G_0$, allowing the measurement of the conductance of single molecule junctions.

In our experimental set-up, the substrate is a 100-200 nm thick Au layer (99.999% purity, Alfa Aesar) thermally evaporated onto a mica substrate and the tip is a freshly cut Au wire 0.25 mm in diameter (99.999% purity Au, Alfa Aesar). A single-axis piezo actuator (Mad City Labs), attached to a micrometer for coarse approach, is used for moving the substrates. A fixed bias of typically 25 mV is applied between the Au wire tip and an Au substrate using a low noise 24 bit data acquisition board (National Instruments PXI-4461) and the resulting current between the tip and sample is converted to a voltage with a current amplifier (Keithley 428-PROG.)

Conductance traces are measured as follows: the substrate is first moved towards the tip in steps of 1 nm until a conductance greater than a few $G_0$ is measured between the tip and substrate. This ensures that the atomic configurations of the gold tip and surface are changed between successive measurements while also allowing the formation of a single atom gold point-contact during each measurement. A voltage ramp is then applied to move the piezo positioner a distance of 5 nm at a rate of 20nm/s while simultaneously reading the current, applied bias voltage and sensor of the Piezo positioner at a 40 kHz sampling frequency. This results in 200 data points per Å of piezo travel. The measured current is divided by the measured bias voltage across the junction, resulting in a conductance versus position trace. All conductance traces that end at the noise limit of the system are saved. Occasionally (less than 1% of the measured traces), the contact does not break within the preset pulling distance due to drift in the system and these traces are not recorded. Before performing measurements on a new solution with target molecules, 1000 traces are measured with a freshly cut gold tip and UV/Ozone cleaned substrate in air to ensure that the gold histogram does not reveal any contaminations in the system. The freshly prepared solution of the molecules (~1 mM in 1,2,4 trichlorobenzene, 99% purity, Sigma-Aldrich) is then added to the substrate and measurements are continued with the same Au tip. Fresh solutions are necessary for each experiment as the diamine solutions oxidize within 24 hours. Although all results presented here are preformed with 1,2,4 trichlorobenzene, we have measured the conductance of molecules in different solvents including dodecane (99% purity, Sigma-Aldrich), phenyl decane (98% purity, Sigma-Aldrich), ethyl benzoate (99% purity, Sigma-Aldrich), butyl phenyl ether (99% purity, Sigma-Aldrich) and dodecanol (99% purity, Sigma-Aldrich). We have found no significant variations in the molecular conductance histograms. However, we cannot use solvents that are low boiling (b.p. below 200 C) or polar solvents.

Conductance traces are analyzed by constructing histograms from all saved conductance traces without any data selection process by binning the conductance data (200 pts/Å) for the entire trace as a function of conductance using a constant bin size. The bin size varies between $10^{-4}$ $G_0$ and $10^{-8}G_0$, depending on conductance region of interest. For all histograms constructed, the counts are divided by the number of thousand traces used to construct histograms so as to be able to compare histograms constructed from different numbers of traces for each molecule.

A conductance histogram constructed from 1000 consecutive traces measured with a UV/Ozone cleaned gold substrate and a freshly cut gold wire in air is shown in Fig. 1 on a log-log scale. Clear peaks are visible around 1, 2, and 3×$G_0$, and a background, primarily due to tunneling, is seen starting below ~$10^{-2}$ $G_0$. When experiments are performed in a solution of the diamine molecules, conductance traces reveal additional steps below $G_0$ due to conduction through a molecule bridging the gap between the two Au point-contacts (inset to Fig. 2). Conductance histograms constructed from all traces measured in the presence of molecules show clear peaks both at multiples of $G_0$, and at a molecule dependent value below $G_0$. This can be seen in Fig. 2, where the histograms of 1,4 diaminobenzene (blue) and 1,4 diaminobutane (red) constructed from 20,000 traces measured consecutively are shown on a log-log scale. Histograms of the two molecules are also shown on a linear scale in Fig. 3 along with a Lorentzian with a functional form: $f(G) = A/((G-G_{peak})^2+B^2)$ fit to the peaks. Here, $G_{peak}$ is the center position of the Lorentzian and gives the most prevalent molecular conductance. The scaled histogram widths defined as $B/G_{peak}$, determined from the fit to the data can be related to the variations in binding geometry and molecule conformations. Both 1,4 diaminobutane and 1,4 diaminobenzene have similar peak widths (~ 0.5), determined from the Lorentzian fits (black trace in Fig. 3a). However, the histogram for 1,4 diaminobutane shows a faint secondary peak (black traces, Fig. 3b) at approximately two times the conductance of the main peak. The region with the two peaks in this histogram can be fit with the sum of two Lorentzians as shown in Fig. 3b. The center of the second is at twice the conductance of the first, indicating that for this alkane, about 10 % of the junctions are formed with two molecules.

The analysis presented so far uses a bin size that is independent of conductance (linear G histogram). It is also possible to generate conductance histograms by binning the log of the measured conductance trace by creating a log G histogram, as has been done recently [27]. Since the data is collected on a linear scale, however, binning the log of G results in decreasing the number of counts at low conductance relative to the counts at high conductance. This is because in a log G histogram, there are a fixed number of bins for each decade in conductance, while in a linear G histogram the number of bins increases by a factor of ten for each decade increase in conductance. To illustrate this and probe the effect on the deduced conductance, log G and linear G histograms are compared in Fig. 4 (red versus blue traces) for conductance measurements of p-terphenyl diamine. The peak in the log G histogram occurs at a conductance value that is 50% larger than the peak in the linear G histogram. Furthermore, these histograms show that the background is not conductance independent as the number of counts at low conductance ($10^{-5}$ G0) is quite a bit higher than at higher conductances ($10^{-2}$ G0). An alternative method to obtain a conductance histogram with low background is to subtract from the linear histogram a histogram obtained from a measurement in solvent alone. The result of such an analysis is shown in Fig. 4 (green trace). A slight shift (10%) in the peak position to higher conductance is also seen here, but it is much less significant than the shift seen in the log G histogram. However, it can also be seen that this approach to background subtraction is not complete. Background counts at the 10-20% level relative to the peak are still seen at low conductance values. Different methods of background subtraction result in different peak conductance values for the same molecule and none provide a clear method to consider the conductance histogram of traces measured with molecules alone. However, our measurements with Amines consistently have a very large fraction of traces with molecular steps (e.g. >95% for this molecule), thus the peak position of the full linear G histogram represents best the conductance value most frequently seen for a gold/terphenyldiamine/gold junction. We use the linear G histograms without background subtraction to obtain all the conductance values reported here.

Part of the data analysed here has been previously published. For the additional data presented here, most of the diamino species were obtained commercially from Sigma-Aldrich, Alfa-Aesar and TCI-America. Two molecules were specifically synthesized for the present work. 4.4' diamino stilbene was prepared by a known method involving the reduction of the dinitro stilbene [41] while bis-(4-aminophenyl)acetylene was prepared by a previously reported method [42].

## 3. Theoretical Methods

The bonding of the amines to the gold electrodes and the associated frontier electronic states were studied using Density Functional Theory (DFT) approaches. The generalized gradient approximation (GGA) as formulated by Perdew, Burke and Ernzerhof (PBE) was used [43]. Model structures were considered to explore different bonding configurations. We primarily focused on the donor-acceptor binding described in the introduction. Two distinct approaches were used. Most of the results reported here were based on using small Au clusters to represent the electrodes. These molecule and finite cluster calculations were done with Jaguar v5.0 [44]. Except where noted, the 6-31g** basis was used for the light atoms and the lacvp** basis was used for Au [45]. Radicals were treated as spin unrestricted. In general, complete geometric relaxation was performed, except where a portion of the Au cluster was frozen to model a small section of the crystal surface. Visualization was done with gOpenMol v2.32 [46].

Certain structures were explored using periodic calculations performed using pseudopotentials [47] and a planewave basis set with Abinit v4.2 [48, 49]. Selected molecules, including cases with adjoined $Au_1$ clusters, were studied in a large enough super cell to isolate the molecules. A 20 Hartree kinetic energy cutoff defined the basis set size, with selective testing up to 30 Hartree cutoff. Bonding was also studied using a slab model to represent the Au electrode surface. The basic supercell consisted of 4 monolayers of Au and 8 monolayers of vacuum, to which adatoms and molecules were added to either one surface or both surfaces. We also considered model junctions in which a diamine molecule spanned the vacuum region, being bonded to both surfaces. This is essentially a slab-molecule superlattice. An Au(111):2x2 surface unit cell was used for most of the calculations. The Brillouin zone was sampled by a 4x4 grid, with selected tests utilizing a 6x6 grid. Some tests were done with an Au(111):3x3 surface unit cell. The calculated lattice parameter of bulk Au (4.182 Å) was used. The surface layer for the flat surface relaxed outward by 2.6%. The primary electrode structure consisted of the addition of an Au adatom to the hcp hollow site. Then the coordinating surface atoms were drawn towards the adatom by about 0.02 Å. The final adatom to surface atom bond length was 2.78 Å. The ammonia and 1,4 diaminobutane calculations on the surface were performed for a structure which maintained one mirror plane orthogonal to the surface and through the molecule.

To assess the accuracy of the GGA calculations for the donor-acceptor bonding, we considered the case of ammonia. Detailed, high quality correlated electron calculations for the radical $AuNH_3$ gave a binding energy for the Au to the ammonia of 0.76 ± 0.06 eV [50]. Our molecular calculations with the GGA (PBE), using a double-zeta basis set, gave 0.70 eV, including a small counterpoise correction. This is somewhat larger than the GGA (PW91) results of Lambropoulos et al. with a similar basis (0.56 eV). Our calculations with a triple-zeta basis set and diffuse functions (lacv3p**++) gave 0.60 eV. Our GGA (PBE) calculations with pseudopotentials and a planewave basis gave 0.61 eV, very close to the GGA (PW91) results of Lambropoulos et al. with a similar approach (0.59 eV). The calculated bond length was 2.28 to 2.29 Å, insensitive to other details. Taken together, these results suggest that GGA (PBE) gives a reasonable account of the donor-acceptor bond with a modest (<0.2 eV) underestimate of the bond energy relative to the accurate result [50].

Based on slab calculations, we found that ammonia binds to the Au(111) surface at the a-top site, with a calculated binding energy of 0.4 eV, similar to a recent calculation [51]. This included a small correction (0.08 eV for the 2x2 unit cell) for additional intermolecular dipole-dipole interactions to obtain the dilute limit. The result is in reasonable agreement with the binding energy inferred from temperature programmed desorption, 0.43 eV in the dilute limit [52]. However, the ammonia was calculated to bind to an Au adatom (hcp hollow site) much more strongly. Based on Au(111):2x2 and Au(111):3x3 unit cell calculations, the binding energy was 0.7 eV. The residual dipole-dople interactions are larger (0.2 eV for the 2x2 unit cell), indicative of a larger induced dipole for the donor-acceptor bond to the adatom. Overall, the stronger binding to a prototypical undercoordinated Au site agrees qualitatively with the broad range of ammonia adsorption energies observed on rough Au surfaces [53]. A similar increase of binding energy was found for isocyanide attached to an adatom on Au(111) [54]. Several small clusters were explored to model the adatom, including $Au_1$ (already discussed), $Au_5$ (the immediate triangle of Au atoms in the surface layer plus one subsurface Au atom), an $Au_{10}$ pyramid and an $Au_{10}$ model for the fcc hollow site (six Au atoms in the surface layer and three from the subsurface layer). In each case, a portion of the cluster is held in a fixed geometry to model a fragment of the Au(111) surface. Double-zeta basis set calculations gave binding energies from 0.70 to 0.84 eV, including counterpoise corrections. In summary, the periodic slab models introduce residual induced dipole-dipole interactions, while the cluster models show size and structure dependent binding energies. However, taken together, the slab and cluster model calculations suggest a fairly consistent result: the ammonia binds to an Au adatom structure with an energy of 0.7-0.8 eV.

## 4. Theoretical Models for Amine Linked Junctions

*4.1. Bonding in Amine Linked Junctions*

To investigate the binding in a junction, we focused on 1,4 diaminobutane. An initial study was performed using a single Au to represent the adatom on each side. We make extensive use of this model below and designate it as an $Au_2$–diamine complex. Our molecular calculations with a triple-zeta basis set and diffuse functions gave a removal energy (breaking both N-Au bonds) of 1.03 eV with N-Au bond lengths of 2.38 Å and C-N-Au bond angles of 119 degrees. For comparison, periodic supercell calculations with pseudopotentials were performed resulting in a removal energy of 0.91 eV with N-Au bond lengths of 2.35 A and C-N-Au bond angles of 118 degrees. The agreement with the molecular calculations is reasonable given the softness of the Au-N bond and differences between the Au pseudopotential and the effective core potential used in the molecular calculations. Molecular calculations with $Au_5$ cluster models for each electrode gave a binding energy of 1.30 eV relative to isolated molecule and clusters, with N-Au bond lengths of 2.34 Å and C-N-Au bond angles of 116 degrees. Junctions were formed in a periodic slab model with the Au(111):2x2 cell (Fig. 5a). The pseudopotential calculations resulted in a binding energy of the molecule in the junction of just 0.5 eV, with N-Au bond lengths of 2.32 Å and C-N-Au bond angles of 119 degrees. Tests with the Au(111):3x3 cell gave a binding energy of 1.5 eV, indicating that the induced dipole repulsion is significant in the junction geometry. Taken together, these results suggest that each donor-acceptor bond is about 0.6-0.8 eV in junctions formed with Au adatoms.

Since the N-Au donor-acceptor bond energy is modest, we sought to characterize the robustness of junctions formed with such links. To probe this, we systematically varied the electrode separation in several junction models. From the energy as a function of distance, the derivative gives the force required to maintain each separation. The inflection point in the energy curve corresponds to the maximum force that the junction can sustain under static conditions. For the 1,4 diaminobutane case bonded to $Au_5$ clusters to model electrodes, the maximum force the junction sustained was 1.0 nN, occurring at a junction elongation about 1.6 Å beyond the optimum [39]. For the periodic slab model illustrated in Fig. 5a, the maximum force sustained by the junction is 0.7 nN (Fig. 5c), about 1.3 Å beyond the optimum. Most of the junction elongation is taken up by stretching the Au-N bond (Fig. 5d), in agreement with a previous calculation [55]. All of the calculations for model junctions were broadly consistent in the qualitative features. Each model presents limitations regarding quantitative conclusions (cluster size or dipole repulsion), but the maximum force fell in the range 0.7 to 1.0 nN. In each case, the Au-N bond ultimately broke.

For comparison, we also considered models for 1,4 diaminobenzene junctions, such as illustrated in Fig. 6a. An $Au_5$ cluster was used to model the electrodes with only the Au adatom free to relax. The calculated binding energy at the optimum separation (double-zeta basis) was 1.11 eV with a N-Au separation of 2.39 Å. The maximum force of 0.6 nN occurred for an elongation of about 1.4 Å (Fig. 6c). About half the elongation is taken up by N-Au stretching with the balance coming from rotation and displacement of the Au adatom about 0.1 Å further from the surface. Both the binding energy and the maximum force are calculated to be somewhat smaller than for the 1,4 diaminobutane. This is consistent with a more spatially directed N lone pair orbital in the latter case.

More realistic models of the junction structure involving more extended Au adstructures would show variations in binding energy. For example, in our detailed study of 1,4 diaminobenzene junctions, different Au motifs led to a range of binding energies, from 0.8 to 1.3 eV [38]. Similarly, the maximum force will vary, but the range calculated here (0.6 – 1.0 nN) should be indicative of the magnitude. For comparison, a measured value for the maximum force sustained by a pure Au-Au single atom point contact was 1.5 nN [56]. This suggests that despite the modest magnitude of the binding energy, the hypothesized donor-acceptor bonding should be tough enough to explain the observed junction formation and the observation of well defined steps in the conductance traces.

*4.2. Conductance through Amine Linked Junctions*

The frontier electronic states reflect the coupling of the N lone pair orbital to the adatom Au s and $d(3z^2-r^2)$ orbitals. The lone pair and the $d(3z^2-r^2)$ orbitals are fully occupied and the s orbital is nominally half occupied. As a consequence the highest occupied orbital has anti-bonding character between the lone pair and the Au orbitals, with the Au s-orbital more prominent than the $d(3z^2-r^2)$ orbital.

The minimal model of the junction formed by a diamine coupled to an undercoordinated Au atom on each electrode is illustrated in Fig. 7a. The single Au atom on each electrode is treated explicitly with coupling to the electrode represented by a self energy in a Green's function approach [57, 58]. The low-bias conductance will follow from the electron transmission through the junction in a Landauer picture. To further simplify the discussion, we focus only on the Au s orbitals and the nearest relevant orbitals on the diamine. So for 1,4 diaminobenzene illustrated in Fig. 7a, the highest occupied molecular orbital (HOMO) is closest, with the HOMO-2 and LUMO+1 coupling to the Au s orbitals with opposite parity. The HOMO-1 and the LUMO are non-bonding pi orbitals with no coupling. The Au s orbitals fall in the gap in the 1,4 diaminobenzene spectrum, but they couple by tunneling through the molecular backbone, in this case via the indicated couplings to the frontier pi orbitals shown. The resulting even and odd orbital combinations for the $Au_2$–1,4 diaminobenzene complex are illustrated in Fig. 8a, the lower one occupied and the upper one empty in the absence of electronic coupling to the electrodes. From the structure of the orbitals on the molecule, one can see that coupling to the HOMO and HOMO-2 dominate in this case. Figure 8b shows the corresponding orbitals for an $Au_2$–1,4 diaminobutane complex. The same gateway combination of N lone pair with Au orbitals leads to tunnel coupling through the sigma states of the alkane backbone. Returning to the example of 1,4 diaminobenzene, the bonding and non-bonding molecular orbitals are shown in Figs. 8c and 8d respectively for the A1g symmetry. This illustrates the role of the Au $d(3z^2-r^2)$ orbitals in the bonding.

The energy splitting between the frontier molecular orbitals of the $Au_2$–diamino complex in Fig. 8a and 8b quantify the tunnel coupling. This measure of the tunnel coupling can be simply related to an estimate of the junction conductance. We simplify the self energy to just a constant imaginary part ($\Gamma$) and restrict the model to two molecular orbitals. Inclusion of another orbital, such as illustrated in Fig. 7a, does not change the essential picture. Then the 4x4 model Hamiltonian together with the embedding self energies for the electrodes is just

$$H_c + \Sigma_L + \Sigma_R = \begin{pmatrix} \varepsilon_s - i\Gamma/2 & \tau & \tau & 0 \\ \tau & \varepsilon_H & 0 & \tau \\ \tau & 0 & \varepsilon_L & -\tau \\ 0 & \tau & -\tau & \varepsilon_s - i\Gamma/2 \end{pmatrix} \quad (1)$$

where the Au s orbital energies are assumed to be the same and the two molecular orbital energies are envisioned as the HOMO and LUMO. The even and odd symmetry of the N lone pair derived frontier states for a diamine are built into the structure of the coupling which is assumed to have constant magnitude $\tau$ (another simplification). The frontier orbital splitting in the complex, uncoupled from the electrodes ($\Gamma=0$) is approximately

$$E_{even} - E_{odd} = 2t \approx \frac{2\tau^2(\varepsilon_L - \varepsilon_H)}{(\varepsilon_L - \varepsilon_s)(\varepsilon_s - \varepsilon_H)} \quad (2)$$

for $\tau$ sufficiently small and $\varepsilon_s$ away from the molecular orbital energies. Use of the notation 2t for this splitting (Figs. 8a,b) expresses an effective electronic coupling, due to tunneling. The ballistic electron transmission through the junction as a function of energy can also be written out analytically for this model from the Green's function. On physical grounds, we can expect that self consistency upon coupling of the molecule adatom complex to the electrodes will keep the adatom Au s orbital energies close to the electrode Fermi energy. Therefore, we evaluate (for zero bias)

$$T(\varepsilon_s) = \frac{\Gamma^2 \tau^4 (\varepsilon_L - \varepsilon_H)^2}{\left((\varepsilon_L - \varepsilon_s)^2 \Gamma^2/4 + 4\tau^4\right)\left((\varepsilon_s - \varepsilon_H)^2 \Gamma^2/4 + 4\tau^4\right)} \approx \frac{4(E_{even} - E_{odd})^2}{\Gamma^2} \quad (3)$$

where the last approximation holds for $\Gamma$ sufficiently large (relative to $16\tau^2/(\varepsilon_s - \varepsilon_H)$ and the complement). The approximate form quantifies the intuitive relationship between the tunnel induced frontier orbital splitting in the complex and transmission through the junction formed by coupling to electrodes.

In Fig. 7b, the transmission versus energy for the five state model (Fig. 7a) is shown for several values of electronic coupling between the Au adatom and the electrode. The energy levels and coupling $\tau$ were chosen to roughly represent the $Au_2$–1,4 diaminobenzene complex. From the DFT calculations, the 1,4 diaminobenzene HOMO to LUMO+1 gap was about 4 eV and the tunneling induced splitting in the complex (2t, in Fig. 8a) was 0.30 eV. The frontier orbitals in the complex are about 1 eV above the 1,4 diaminobenzene HOMO, placing the Fermi energy in this model calculation (arrow). Each 1,4 diaminobenzene orbital in the model gives a transmission resonance. In addition, for small electronic coupling to the electrode (e.g. $\Gamma=1$ eV), there is a residual transmission resonance associated with the Au adatom s orbitals. However, for larger coupling, the Au adatom orbital acts as a gateway to the molecule, but no longer supports an identifiable resonance. By comparison to full calculations of the transmission through 1,4 diaminobenzene junctions [38, 55], $\Gamma=2$-3 eV is representative. The low bias conductance of the junction is dominated by the HOMO orbital. The alignment of the electrode Fermi energy on the flank of this resonance results in conductance less than $G_0$. When DFT orbital energies are used in this calculation, the conductance is overestimated because the HOMO resonance is too close to the Fermi energy [38].

The calculations for a complex with a single Au atom representing the contact to each electrode can be readily done for all the molecules studied. The DFT calculations for an $Au_2$–diamino complex provided the effective tunneling coupling through each of the molecules used to form junctions. The chemical potential for the $Au_2$–diamino complexes (average of HOMO and LUMO energies, such as illustrated in Figs. 8a,b) varied quite modestly across all the species studied, a range of about 0.5 eV. In terms of the model in Fig. 7a, the self consistent Au s orbital energy varied modestly. Therefore, it is reasonable to assume that the electronic coupling Γ will be insensitive to the details of the diamino species. Then using Eq. (3), the tunnel coupling $E_{even} - E_{odd} = 2t$ calculated explicitly for each complex allows us to study trends between different species and for different configurations of the molecules.

For example, varying the N-Au bond length in the 1,4 diaminobutane case by ±0.15 Å from the optimum increases the energy by about 0.1 eV, but causes a variation in the tunneling probability ($4t^2$) of less than 15%. On the other hand, varying the C-N-Au angle by ±15 degrees from the optimum also increases the energy by about 0.1 eV, but results in about 35% variation in tunneling probability. Sensitivity to bond angle is indicative of the directionality of the N lone pair orbital. We envision that each new junction formed in the experiment is constrained by the local arrangement of the Au electrodes so that the local bonding of the 1,4 butantediamine to form a junction will be suboptimal and sample a range of N-Au bond lengths and C-N-Au bond angles. For comparison, the calculated relative energy of the junction at the maximum force is ~0.4 eV (Figs. 5b, 6b). These results illustrate that the corresponding range of conductance will be modest, but contribute to the measured width of the histograms.

For the junction model shown in Fig. 5a, the tunneling coupling was extracted from the frontier orbital band width in the periodic slab calculation. As shown in Fig. 5e, the tunneling probability only varies modestly over the range of junction elongation where the 1,4 diaminobutane is bound in the junction. Similar results were found for 1,4 diaminobenzene. The tunnel coupling for the $Au_5$ model electrodes shown in Fig. 6a was calculated from the average of the three pairs of frontier orbitals. As a function of elongation (Fig. 6e), the tunneling probability is more constant than the 1,4 diaminobutane case. The N lone pair in the 1,4 diaminobenzene case is coupled into the pi system on the ring with the HOMO having a nodal plane between the N and the C (Fig. 7a). As the Au-N-C angle increases, the tunnel coupling responds to two competing factors: the intrinsic angular dependence of the N lone pair drives the coupling down, but the Au s orbital is further from the nodal plane, compensating.

We have used the tunnel coupling calculated from the $Au_2$–diamino complexes to assess trends within several molecular families [34-37]. The trends compared well with the measured conductance and helped in the interpretation. Here, we have extended these results and show comparisons across molecular families. To do so, the electronic coupling of the adatom to the electrode is assumed to be the same, independent of the molecular backbone. The self consistent DFT calculation for the $Au_2$–diamino complex includes all the details of the bonding, charge rearrangments and relative alignment of the frontier tunneling states to the molecular backbone states, in this simplified picture. Then, with reference to Eq. (3), the tunnel coupling is explicitly calculated and the coupling to the electrode (Γ) is scaled away by using the 1,4 diaminobenzene as a reference case. This does not solve the inherent problems with using DFT orbital energies [38], but it does enable broad comparisons for amine linked molecules.

## 5. Trends Observed in Single Molecule Conductance

### 5.1. Length and Conjugation

The conductance of single molecule junctions measured with amine links demonstrated very clearly the trend expected in the case of non-resonant tunneling through a molecular backbone [34]. In Fig. 9a the conductance histograms measured for a series of alkanes with amine end-groups that progresses from 1,2-ethylenediamine to 1,12-dodecanediamine are shown, extending to longer molecules. These histograms display a peak and the conductance of these molecules is determined from the center position of a Lorentzian fit to the peak. As shown in Fig. 9b, the conductance decreases exponentially with increasing length, with a decay constant of 0.76 ± 0.01 per Å or 0.97 ± 0.02 per methylene group. For comparison, we also plot our previously measured conductance values for a series of oligophenyls with 1, 2 and 3 phenyl rings [35]. The conductance versus molecule length fits an exponential form with a decay constant of 0.43 ± 0.02 per Å or 1.8 ± 0.1 per phenyl ring, in agreement with a previous measurement [23]. A third example consists of a series of acenes with 1, 2, and 3 benzene rings. The 2,6 diaminonaphthalene and 2,6 diaminoanthracene probe conductance along the length of these fused ring aromatic molecules. Here, we found a deviation from an exponential decay (Fig. 9b) which we understood to be due to a significant change in the HOMO-LUMO gap with increasing length in the acenes [37].

For comparison the calculated tunneling probability is superposed in Fig. 9b with the scale adjusted so that experiment and theory coincide for 1,4 diaminobenzene. The calculated tunneling probability for the alkane series and the acene series, relative to benzene, agree very well with the magnitude of the measured conductance. There is some systematic underestimate of the short alkane conductance (about 25%). Fitting the alkane series to an exponential yields a decay constant of 0.93 ± 0.01 per methylene group, in good agreement with experiment. However, the fit for the oligophenyl series gives a decay constant of 1.50 ± 0.03 per phenyl unit. This difference in slope between the calculation and the experiment is quite apparent in Fig. 9b. Since the calculated decay constant is smaller than experiment, it suggests that the electrode Fermi energy is too close to the diamino HOMO in this case [59]. This is consistent with the more detailed analysis of 1,4 diaminobenzene [38]. Finally, the calculations for the acene series were extended out to N=5 (2,6 diaminopentacene), highlighting the non-exponential dependence on length for the acene case (Fig. 9b).

*5.2. Short "Molecular Wires" and Conformation*

Using amine links, we have been able to measure the conductance of several short oligomers of "molecular wires," conjugated organic molecules with useful conducting characteristics in polymeric form. We measure and compare the conductance of three molecular wires: 4,4' diamino azo benzene (OPA), 4,4' diamino stilbene (OPV) and bis-(4-aminophenyl)acetylene (OPE). Conductance histograms for these three molecules are shown in Fig. 10. The conductance value of these three molecules varies systematically, from the position of the peak in the histograms, with the Azo having the highest conductance. Interestingly, the high conductance end of the histogram for the ethynyl molecule extends beyond that of the azo possibly due to inclusion of a second molecule in some junctions or the formation of an Au-C link in some junctions [60]. Specifically, the most probable conductance increases from the ethynyl ($5.7\times10^{-4}$ $G_0$) to the vinyl ($6.1\times10^{-4}$ $G_0$) to the azo ($7.4\times10^{-4}$ $G_0$). Our finding that the ethynyl case has lower conductance than the vinyl case agrees with ensemble measurements for longer oligomers (three phenyls instead of two here). However, the ensemble measurements, performed for SAMs on crossed wire junctions [25] and in-wire junctions [26]), show a larger ratio between the OPV and OPE conductance (about 3x). In our case, the relatively small changes in conductance are consistent with the differences in N-N distance (assuming all three species to be approximately flat) and an exponential decay constant of 0.4 per Å. If we compare our OPV (N-N distance 12.3A) to a comparable length alkane (C9 in our case, N-N distance 12.8 A, most probable conductance $6.2\times10^{-6}$ $G_0$), we find a ratio of about 100 times. In the ensemble case, they compared to C12, finding a ratio for OPV of 46 times. The DFT calculations for the tunnel coupling show a trend from Azo to Vinyl to Ethynyl that is very close to experiment. However, based on scaling to 1,4 diaminobenzene, the predicted conductance is about 2.5 times the measured values. Therefore the predicted conductance ratio for OPV to C9 is about 300.

By considering a series of substituted biphenyls, we measured the explicit dependence of single molecule conductance on the average conformation of the molecule, the twist angle at the central C-C bond of biphenyl [35]. Because the twisting degree of freedom is quite soft, it raises further interesting questions concerning the role of molecular conformation in the measured single molecule conductance. A striking feature in the histograms for the three wires in Fig. 10 is the systematic trend in the widths of the molecular conductance peaks. The azo molecule has the narrowest width, while the ethynyl has the broadest peak. Based on the Lorentzian fits, the widths relative to the most probable conductance increase from azo (0.8) to vinyl (1.1) to ethynyl (1.6). This suggests that these differences in the interior linkage are affecting the different configurations in which these single molecule junctions can form.

For a specific junction (one conductance trace in our experiment), the measured conductance on a single step represents a thermal average over accessible configurations for that particular junction. With regard to twisting of the biphenyl, this average gives a result very close to the static result for the most probable twist angle. However, we measure a large number of individual junctions, each one based on breaking a new Au point contact in solution. Therefore, when considering the histograms of conductance, it is crucial to consider three other sources of variation. First, the local environment of the local Au atom on the electrode that is bonded to the amine varies. Second, in order for the diamino species to form the two bonds required to make circuit, there will inevitably be accommodation to the steric constraints of the specific local shape of the two Au electrodes. This results in some deformation of the N-Au bond length and the C-N-Au bond angle which affects the tunneling probability, as discussed in Sect. 4.2. We have shown that these two effects together gives rise to a spread in the junction conductance for 1,4 diaminobenzene that is similar to the width of the measured histogram [38].

For more complex molecules, such as the biphenyl, there is a third conformational effect. The constraints of junction formation can be accommodated through the low energy internal distortions, such as the internal twist angle. Interestingly, the conductance histograms for these biphenyls exhibit a conductance peak width, as determined from a Lorentzian fit to the data, that is broader than the width for 1,4 diaminobenzene. The peak width for 4,4'' diaminoterphenyl is even broader [35]. This suggests that the addition of soft internal degrees of freedom has a significant effect on the distribution of junction conductance values. The broader conductance distribution also may be attributable, in part, to increased length, giving access to more configurations. However, the histograms for the 1,N diaminoalkanes show peak widths that are similar, independent of length, and close to the width of the histogram peak for 1,4 diaminobenzene. This suggests that molecule length is not the determining factor, although it does complicate comparisons. Within the set of substituted biphenyls, the spring constant associated with the twist angle does vary. However, the substituents may also have a direct influence on the N-Au, confusing comparisons. The molecular wires presented in Fig. 10 are all of similar length and have very similar bonding to the Au. They present a good example for further study.

In order to explore the role of molecular configuration, we performed a series of constrained calculations for each species, probing the energy and tunneling probability as a function of torsional distortion. As illustrated in Fig. 11, we systematically rotate one Au-NH$_2$ group relative to the other. Both Au-NH$_2$ groups were frozen to impose the torsional constraint, but the rest of the molecule was allowed to relax. Physically, these constraints are envisioned to be imposed by other aspects of the electrode structure. The total energy change is mapped in Fig. 11a. A rotation of 180 degrees corresponds to a change from the configuration where the N-Au bonds are on opposite sides of the backbone to the configuration where they are on the same side, the latter being at a slightly higher calculated energy (5 meV). The phenyl rings relax somewhat against the constraint of the NH$_2$ groups (less than 20 degrees). However, most of the imposed rotation is taken up at the inner linkage between the two rings, the ethynyl, vinyl or azo group. Interestingly, the overall energy versus angle is similar for all three groups. However, the changes in tunneling probability with rotation are systematically different (Fig. 11b). For the ethynyl link, there are no additional structural degrees of freedom. However, with rotation, the pi states of the phenyl rings mix with the orthogonal pi states on the ethynyl group. As illustrated in the inset, this results in the pi molecular orbitals from the two halves of the molecule being nearly orthogonal on the ethynyl for 90 degree rotation. In contrast, in the vinyl case, the vinyl group distorts in response to the torsional constraint, maintaining much stronger electronic coupling through the pi system. As a consequence, the tunneling probability through the molecule is much less affected by rotational distortions. The azo case is even less affected because the central N-N pulls in more weight from the frontier pi orbitals, stabilizing the coupling further. Thus the widths seen in the histograms correspond well to the changes in tunnel coupling expected due to formation of junctions with a distribution of different torsional angles.

*5.3. Comparison Across Molecular Families*

We now have well defined measured values of conductance for amine linked molecules from five distinct families. In this section we give a summary comparison across those families. The five families consist of:

(i) 1,4 diaminobenzene and 11 cases altered by substuents replacing one or more hydrogens on the phenyl ring [36].

(ii) 1,N diaminoalkanes with N=2 to 12 methylene units from Sect. VA.

(iii) 4,4' diaminobiphenyl and 9 cases altered by substituents at different points on the phenyl rings This includes data [35] and measurements for 3,5,3',5' Tetramethyl 4,4' diaminobiphenyl ($G = 1.56 \times 10^{-3}$ $G_0$), 3,3' dimethyl 4,4' diaminobiphenyl ($G = 1.32 \times 10^{-3}$ $G_0$), and 3,8 diamino 6 phenyl phenanthridine ($G = 1.48 \times 10^{-3}$ $G_0$).

(iv) The molecular wires measurements from Sect. V.B, to which we add 4,4'' diaminoterphenyl [35].

(v) Acenes (naphthalene, anthracene) with the amines creating a conduction path both perpendicular and parallel to the long axis of the molecule (total of four cases) [37].

This organization places the three molecules from the oligophenyl series (benzene, biphenyl and terphenyl) in separate categories; the oligophenyl case was specifically mapped in Fig. 9. All together the data set includes 41 distinct molecules.

To put experiment and theory on a common scale, we use 1,4 diaminobenzene as the reference (measured $G_{benzene} = 6.4 \times 10^{-3}$ $G_0$). The measured conductance of the other 40 molecules are scaled $G_{benzene}$. The calculated tunneling has been scaled by taking the ratio of $4t^2$ to the value calculated for 1,4 diaminobenzene ($2t_{benzene} = 0.30$ eV).

A correlation plot is shown in Fig. 12, with each of the five molecular families displayed using a distinct symbol. The full range of the measured data spans five orders of magnitude. Overall, the correlation between the measured conductance and the calculated tunneling probability is remarkable. Despite the use of DFT orbitals, the relative alignment of the frontier orbitals to the molecular backbone orbitals captures the main trends across all five families. Closer examination shows that both the biphenyl series and the molecular wire series fall distinctly above the one-to-one trend line. As noted above, for the molecular wire series, the predicted conductance is systematically off by a factor of about 2.5. For the biphenyl series, the deviation averages about 30%.

The hidden variable in Fig. 12 is molecular length. Referring back to Fig. 9, the calculated tunnel coupling for the alkanes shows an exponential decay constant close to experiment. In terms of the level alignment, the Fermi energy is relatively close to mid-gap where the decay constant is insensitive to errors in level alignment [59]. However, the calculated length dependence for the oligophenyl series gives a decay constant that is about 15% smaller than experiment. This is consistent with the level alignment in the DFT calculations placing the Fermi level too close to the HOMO [38]. Although the error in decay constant is small, the relative error in conductance grows rapidly (30% for biphenyl and 200% for terphenyl). The errors that trace to level alignment do not completely scale out of the conductance problem unless the Fermi energy is close to mid-gap (e.g. Eq. 2). Returning to the trends in Fig. 12, small errors in the decay constant for the oligophenyl series map to the systematic shifts in the trend lines placing the molecular wire series and the biphenyl series above the one-to-one line.

## 6. Concluding Remarks

We have reliably and reproducibly measured the conductance in single molecule junctions using Au-Amine link chemistry across several distinct types of molecules and for over 40 different molecules. This enables the study of systematic trends in single molecule conductance within specific types and across different families of molecules. The unifying hypothesis for the reproducible junction formation is the selective donor-acceptor bonding between the N lone pair orbital on the amine groups and undercoordinated Au atoms on each electrode. While DFT based calculations show that this bond is not too strong (estimated 0.6-0.8 eV), the junctions so formed are estimated to sustain forces up to about 1 nN. This, together with the modest dependence of the local electronic coupling on the amine-Au local structure, naturally explains the observation of the well defined steps in the measured conductance versus elongation traces. The extension of the calculated tunnel coupling to all the molecules studied showed good agreement with experiment across the molecular families studied, further supporting the broad applicability of the donor-acceptor bonding motif.

Overall, we have found that the selective binding of the amines to undercoordinated gold atoms on the electrodes provides a bottom-up solution to forming single metal-molecule-metal junctions with a narrow distribution of electronic properties, enabling a detailed exploration of structure-conductance relations for a range of molecules.

**Acknowledgements:** This work was supported primarily by the Nanoscale Science and Engineering Initiative of the National Science Foundation (NSF) under NSF award number CHE-0641523 and by the New York State Office of Science, Technology, and Academic Research (NYSTAR). This work was supported in part by the US Department of Energy, Office of Basic Energy Sciences, under contract number DE-AC02-98CH10886 and by the MRSEC Program of the NSF under Award number DMR-0213574.

**Figures**

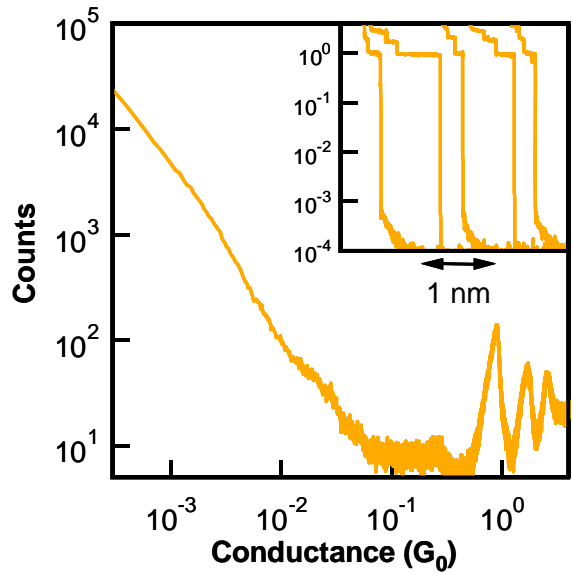

Figure 1. Histogram for reference Au-Au junction formation in air. Log counts versus log of conductance; bins used for histogram linear.

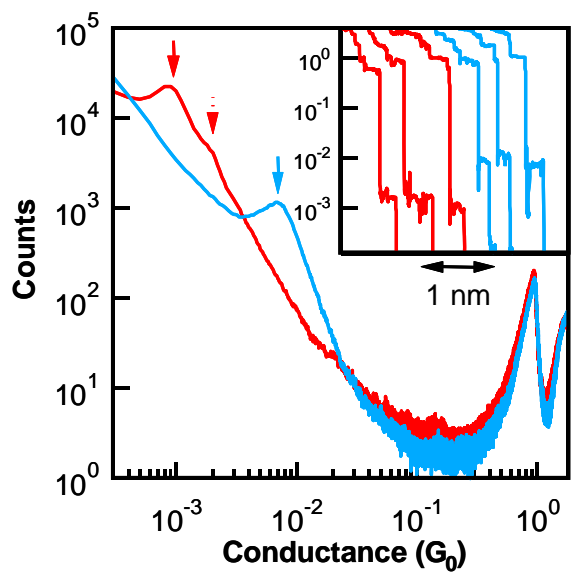

Figure 2. Comparison of histograms for 1,4 diaminobutane (red) and 1,4 diaminobenzene (blue). Arrows indicate peak positions, with a faint secondary peak for the 1,4 diaminobutane case. Inset: individual sample traces.

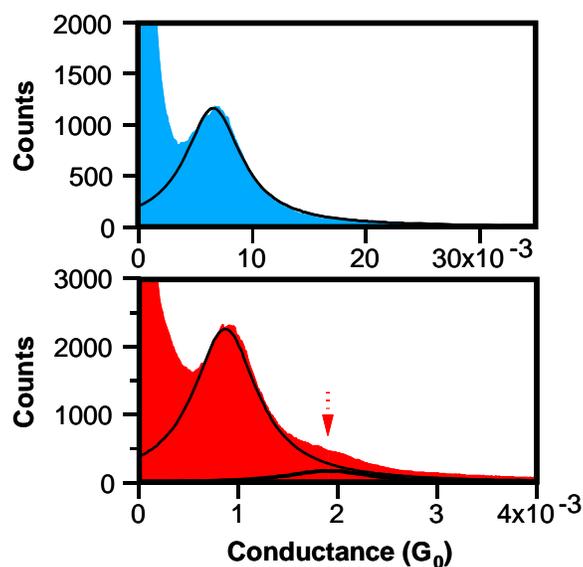

Figure 3. Same data as Figure 2, on a linear scale, illustrating the fitting of the peak by a Lorentz line shape. (a) 1,4 diaminobenzene fit by a single peak. (b) 1,4 diaminobutane fit by two peaks. The shoulder that seems apparent on the log scale in Figure 2 is fit by a second peak with about twice the width and 10% relative amplitude at twice the conductance value.

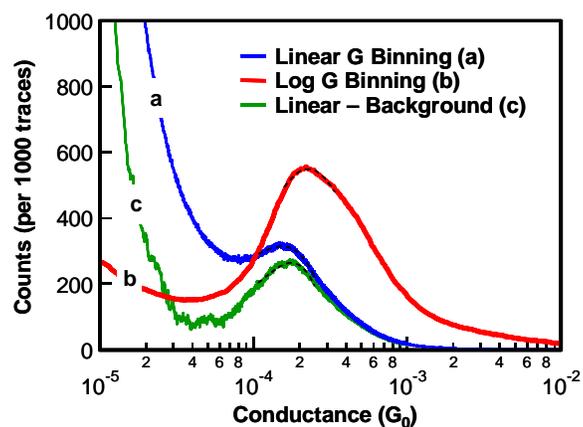

Figure 4. Conductance histograms determined from 24000 conductance traces measured in the presence of 4,4''diamino p-terphenyl using three different methods: (a) Blue: binning Linear conductance traces using a bin size of $10^{-7}G_0$; (b) Red: binning Log of conductance using a bin size of 100 per decade; (c) Green: Subtracting a linear histogram of the solvent alone from blue trace.

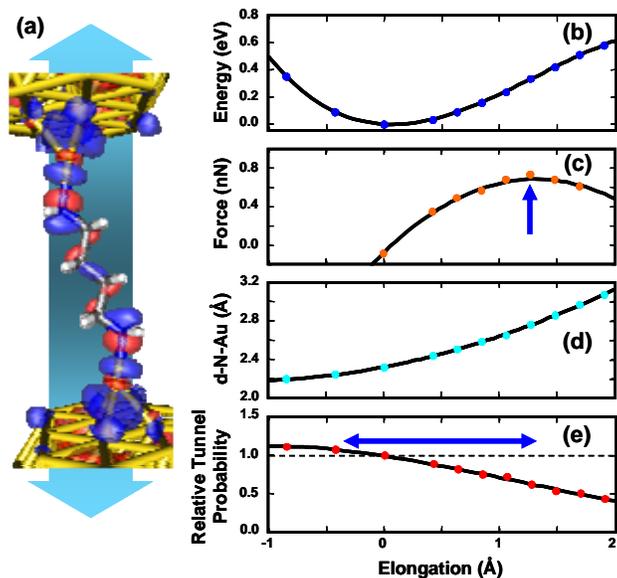

Figure 5. (a) Structure of the 1,4 diaminobutane junction formed by bonding to an Au adatom on each electrode. The plot isolates a single molecule from a 2x2 periodic array. The superposed isosurface plot illustrates the frontier orbital responsible for tunneling. As a function of junction elongation, the relative energy (b), applied force (arrow indicates maximum) (c), N-Au distance (d) and relative tunnel probability (e) are plotted. The arrow in (e) indicates a plausible range for a conductance step.

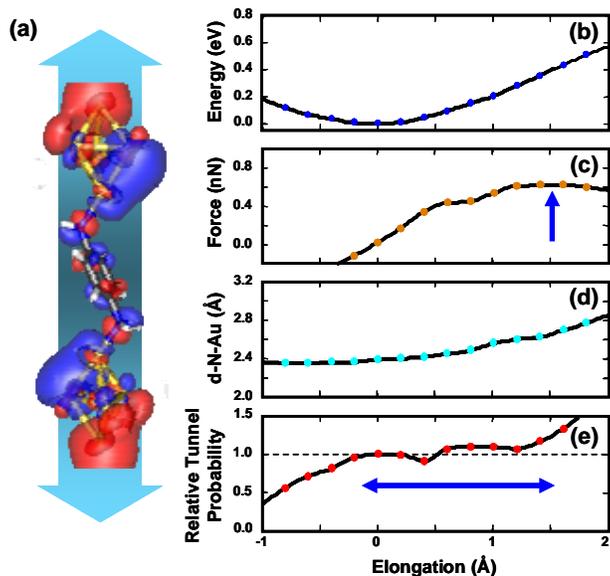

Figure 6. (a) Structure of the 1,4 diaminobenzene junction formed by bonding to an Au adatom on each electrode. The each adatom plus electrode is modeled by an $Au_5$ cluster. The superposed isosurface plot illustrates a frontier orbital responsible for tunneling. As a function of junction elongation, the relative energy (b), applied force (arrow indicates maximum) (c), N-Au distance (d) and relative tunnel probability (e) are plotted. The arrow in (e) indicates a plausible range for a conductance step.

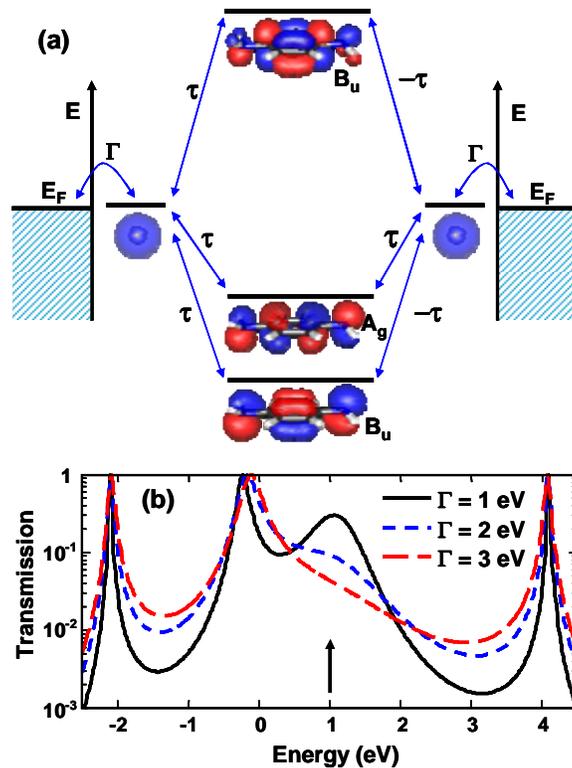

Figure 7. (a) Model for the frontier electronic states that control conductance through a 1,4 diaminobenzene junction bonded to an Au adatom on each electrode. Isosurface plots illustrate the frontier molecular states and the Au s orbital. (b) Calculated transmission versus energy (zero bias) for the five level model illustrated in (a) for three values of coupling to the reservoirs.

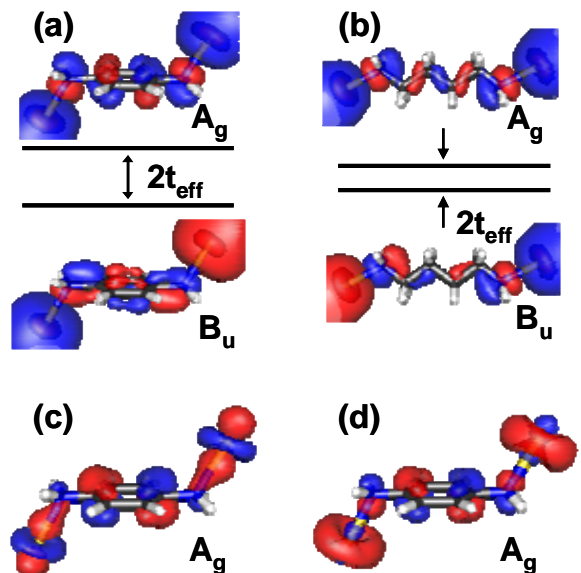

Figure 8. (a) The HOMO and LUMO orbitals of the $Au_2$–1,4 diaminobenzene complex (isosurface plots) with a schematic indication of the effective coupling due to tunneling. (b) The HOMO and LUMO orbitals of the $Au_2$–1,4 diaminobutanecomplex (isosurface plots) with a schematic indication of the effective coupling due to tunneling. (c) and (d) Isosurface plots of symmetric bonding and non-bonding molecular orbitals of the $Au_2$–1,4 diaminobenzene complex illustrating the role of the Au $d(3z^2-r^2)$ orbital.

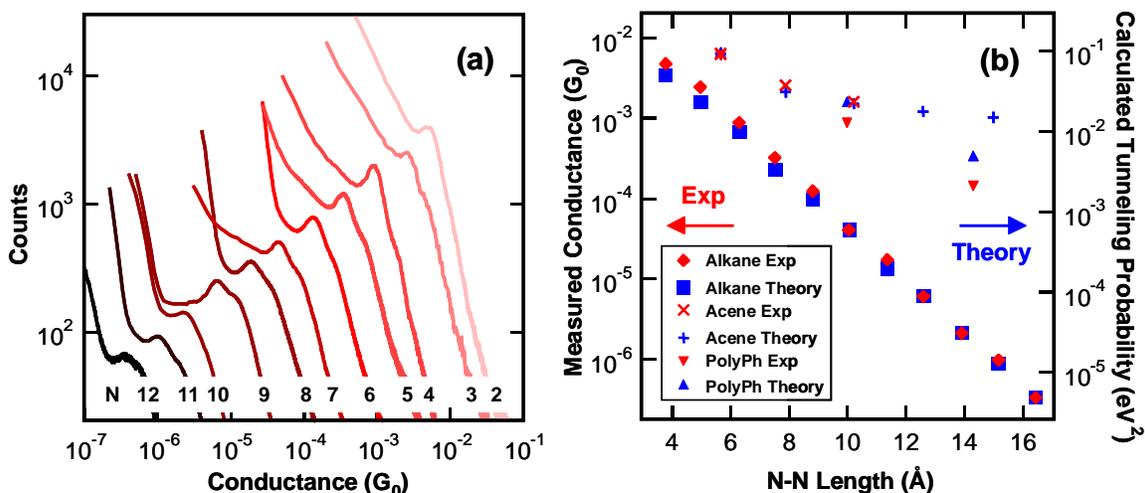

Figure 9. (a) Conductance histograms for a series of 1,N diaminoalkanes, N=2 to 12. (b) Most probable conductance for the alkane series plotted as a function of the N-N separation. Also shown are the same data for diamino polyphenyls (N=1 to 3) and diamino polyacenes (N=1 to 3). For comparison, the calculated tunnel probability is plotted for the same three families, but with polyacenes extended to N=5. The right hand scale is chosen so the theory and experiment coincide for 1,4 diaminobenzene.

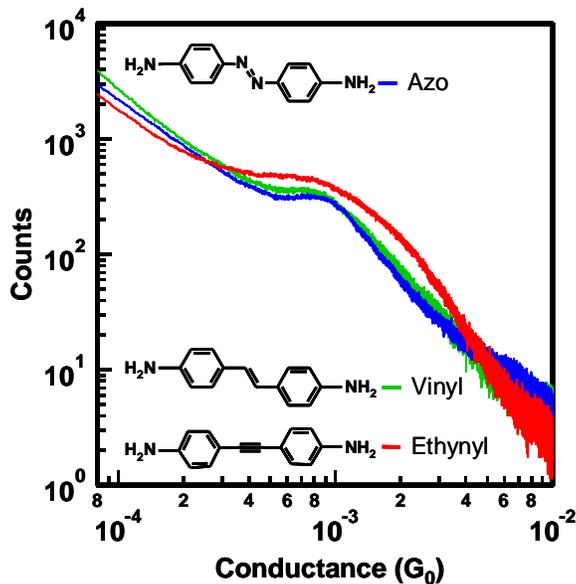

Figure 10. Conductance histograms for three variations on short molecular wires designated by the insets (bis-(4-aminophenyl)acetylene, Ethynyl; 4,4' diaminostilbene, Vinyl; 4,4' diaminoazobenzene, Azo).

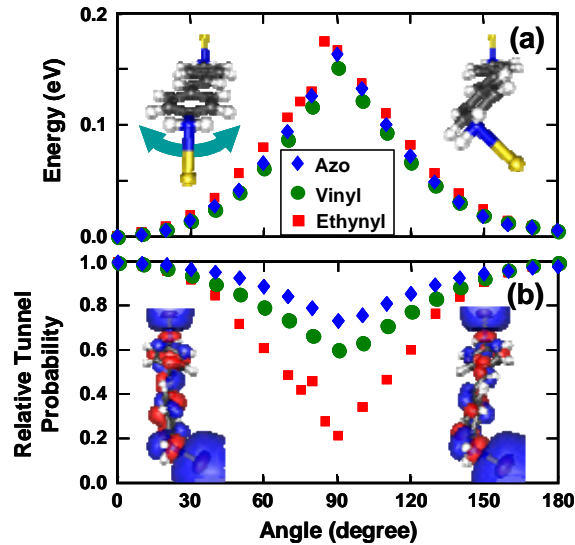

Figure 11. Calculations for the three short molecular wires from Figure 10 with each amino group bonded to $Au_1$, probing the relative rotation of the two link groups. (a) energy versus relative rotation angle. Insets illustrate the initial rotation for the vinyl case. (b) Relative tunnel probability versus rotation angle. Insets illustrate one of the tunnel coupled frontier orbitals with an isosurface plot for the ethynyl (left) and vinyl (right) cases, both for 90 degree rotation.

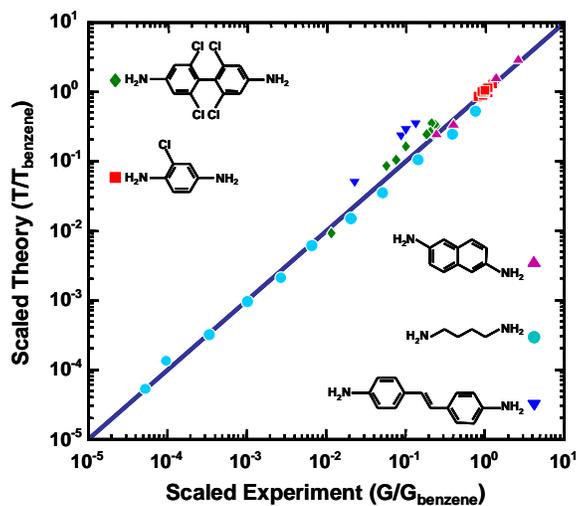

Figure 12. Comparison of scaled theory to scaled experiment with the one-to-one line indicated, on a log-log plot for 41 molecules grouped into five families. 1,4 diaminobenzene is the reference case for scaling. The insets provide a key and illustrates an example from each family.